\documentclass[a4paper,12pt]{article}
\usepackage[utf8]{inputenc}
\usepackage{amsmath,amsfonts,dsfont}
\usepackage[dvipsnames]{xcolor}
\usepackage{authblk}
\usepackage{amssymb}
\usepackage{tikz}
\usepackage[T1]{fontenc}
\usepackage{color}
\usepackage{latexsym}
\usepackage{amsbsy}
\usepackage{amstext}
\usepackage{graphicx}
\usepackage{amsmath}
\usepackage{amssymb}
\usepackage{tikz}
\usetikzlibrary{intersections, angles, quotes}
\usetikzlibrary{positioning}
\usepackage{tikz-cd}
\usetikzlibrary{arrows,decorations.markings}
\usetikzlibrary{quotes, calc}
\usepackage{mathtools}
\usepackage{amsthm}
\usepackage{amstext}
\usepackage{amssymb}
\usepackage{graphicx}
\usepackage{esint}
\usepackage{times}
\usepackage{color}
\usepackage{fancyhdr}
\usepackage{amsfonts,amsthm,amsmath,amssymb,graphicx,hyperref,youngtab}
\usepackage{amsmath,amsfonts,dsfont}
\usepackage{amssymb}
\usepackage{tikz}
\usepackage[T1]{fontenc}
\usepackage{latexsym}
\usepackage{amsbsy}
\usepackage{amstext}
\usepackage{graphicx}
\usepackage{amsmath}
\usepackage{amssymb}
\usepackage{tikz}
\usepackage{mathtools}
\usepackage{amsthm}
\usepackage{amstext}
\usepackage{amssymb}
\usepackage{graphicx}
\usepackage{esint}
\usepackage{times}
\usepackage{fancyhdr}
\newcommand{\cA}{\mathcal A}

\newcommand{\cC}{\mathcal C}
\newcommand{\cD}{\mathcal D}

\newcommand{\cF}{\mathcal F}

\newcommand{\cL}{\mathcal L}
\newcommand{\cM}{\mathcal M}
\newcommand{\cN}{\mathcal N}

\newcommand{\cZ}{\mathcal Z}

\newcommand{\be}{\begin{equation}}
\newcommand{\ee}{\end{equation}}
\newcommand{\bea}{\begin{eqnarray}}
\newcommand{\eea}{\end{eqnarray}}

\DeclareMathOperator{\Arctanh}{Arctanh}

\title{Quiver Asymptotics and Amoeba: Instantons on Toric Divisors of Calabi-Yau Threefolds}
\author{Ali Zahabi}
\affil{Institut de Mathématiques de Bourgogne, UMR 5584, CNRS, Université Bourgogne Franche-Comté, 21000, Dijon, France}
\date{}
\begin{document}
\maketitle
\begin{abstract}
{The BPS bound states of D4-D2-D0 branes on the non-compact divisors of Calabi-Yau threefolds and the instantons in the dual quiver gauge theories are previously studied using two-dimensional crystal melting model and dimer model, \cite{Nish}. Using the tropical geometry associated with the toric quiver, we study the asymptotic of the quiver gauge theory to compute some of their thermodynamic observables and extract the phase structure. We obtain that the thermodynamic observables such as free energy, entropy and growth rate are explicitly obtained from the limit shape of the crystal model, the boundary of the Amoeba and its Harnack curve characterization. Furthermore, we observe that there is a Hagedorn phase transition in the instanton sector inferred from the Gumbel distribution of the fluctuations in the crystal model. We present explicit computations of the results in some concrete examples of $\mathbb{C}^3$, conifold, local $\mathbb{P}^1\times \mathbb{P}^1$ and local $\mathbb{P}^2$ quivers.}
\end{abstract}
\tableofcontents
\section{Introduction}
The D-brane configurations and their corresponding BPS states in quiver gauge theories have been a fruitful field of study in non-perturbative effects in gauge/string duality. They have been studied extensively with a variety of plausible methods, including some integrable models. In particular, the D6-D2-D0 brane bound states in IIA string theory on toric Calabi-Yau threefold singularities and their dual, the 4d, $\cN=2$ toric quiver gauge theories are studied via 3d crystal melting model and toroidal dimer model \cite{Ok-Re-Va,Ya-Oo,Ya-Oo2}.

In a similar approach, the D4-D2-D0 bound states are obtained by replacing the D6-brane with a D4-brane on a non-compact toric divisor of the Calabi-Yau threefolds. Similar to the parent D6 theory, in which the D4 theory is embedded,  the low energy effective theory or the dual quiver gauge theory of the D4 brane is characterized by the toric geometry data of the Calabi-Yau, subject to additional constraints imposed by the embedding \cite{Nish}.
The quiver gauge theory in the world-volume of the D4-brane is the topologically twisted $\cN=4$ supersymmetric gauge theory, called \textit{Vafa-Witten theory}. In fact, the low energy effective field theory on the bound states of $N$ D4-branes wrapped on the four-cycles with D2's wrapping two-cycles and $k$ D0's branes is equivalent to the $k$-instantons in the Vafa-Witten theory. The BPS index of the quiver which is the degeneracy of D4-D2-D0 bound states is the Euler characteristics of the instanton moduli space.
  
Regarding the BPS states and instantons of the quiver gauge theory, all the relevant data for the construction and counting of them are encoded in a two-dimensional crystal model living on the facets of the three-dimensional crystal model. The 2d crystal model is constructed from the representation of the moduli space of vacua of D4 theory, called modules of path algebra.

We continue the study of a D4 brane on a non-compact divisor and D2-D0 BPS bound states to that, using the crystal melting model, which is initiated in \cite{Nish}. Our goal is to study the asymptotic of this theory and compute the thermodynamics observables and extract the possible phase structure of the system. We revisit the well-known instanton counting of D4-D2-D0 bound states in the class of toric divisors and compute the entropy of the gas of instantons. In other words, we study the Vafa-Witten theory and compute the asymptotics of Euler characteristic of the moduli space of the instantons in the limit of a large number of instantons.

In general, the asymptotic analysis is based on the generating function of the BPS states and it is performed by the analysis of the poles of the generating functions. However, there are only few known explicit generating functions for the BPS states, and thus one has to deal with other alternative plausible approaches. In our previous works \cite{Zah,za-iso}, we introduced the applications of the statistical dimer model and associated Mahler measure, as well as the hyperbolic geometry, in the asymptotic analysis of the BPS sector and the computations of the BPS growth rate, entropy and free energy of the toric quivers and the particular class of isoradial ones. The asymptotic analysis of the original BPS counting in the 3d crystal melting model is based on the limit shape of the crystal, i.e. the mirror curve of the Calabi-Yau or the Ronkin function and its closed cousin the Mahler measure of the statistical dimer model \cite{Za}.

In this paper, we study the thermodynamics and critical phenomena of the quiver theories on the toric divisors. The new asymptotic techniques proposed in this study are based on the tropical geometry of the Calabi-Yau threefolds.  In analogy to the 3d crystal model, in this study we use the limit shape of the 2d crystal model, which is obtained from the Amoeba associated with the Newton polynomial of the dimer model, and using a large deviation problem we derive explicit results for the thermodynamics observables such as the free energy and the entropy of the instanton sector. In brief, the entropy density is directly computed by the Legendre transform of the limit shape. Using the Legendre duality between entropy and free energy we obtain the free energy as the area of the Amoeba. Then, by using the tropical geometry and Harnack curve characterization of the Amoebas of the toric quivers, we obtain a simple explicit result for the free energy of the quiver given by the area of the Newton polygon. Furthermore, the instanton growth rates are computed either via the extremization of the entropy function or the saddle-point technique. Eventually, we observe that the total free energy and total BPS growth rate summed over all the divisors are dependent of K\"{a}hler moduli.

Adopting and generalizing some results in number theory, namely the Erd\"{o}s-Lehner results \cite{er-le} about the asymptotic analysis of the restricted integer partitions, we extract the phase structure of the instanton sector. Roughly speaking, the statistical distribution of the fluctuations around the boundaries of a 2d crystal model captures the phase structure of the system. The asymmetric distribution of the fluctuations, known as Gumbel distribution, determines a Hagedorn phase transition in the instanton sector. This phase transition can be explained and interpreted from the limit shape formation and condensation of instantons.

Besides the nice geometrical interpretation and explicit new results that our geometric method for asymptotic analysis of quivers brings into the picture, the main advantage and importance of the method is that it can provide explicit asymptotic results even for the quivers/geometries that their BPS generating function is not yet available. In the examples in which the generating function is available we compare our results obtained from the geometric method with the results obtained from standard asymptotic analysis of the generating function and we find good agreements.

The rest of the paper is organized as follows. In chapter two, we review the physics background of the D-brane bound states and the quiver gauge theories, and the two-dimensional crystal melting model. In chapter three, the main results of the paper in the asymptotic analysis of the quiver gauge theory is explained. In chapter four, our method and results are implemented in some concrete examples.

\section{Quiver BPS states, instanton counting and crystal melting}
In this part, we describe the main problem of this study, the asymptotic counting of the BPS states of the quiver gauge theory, and the instantons of the topologically twisted $\cN=4$ super Yang-Mills theory, the low energy effective theory in the world-volume of the D4 brane. The BPS states/instantons are the D2-D0 brane bound states to the D4 brane. 
After a brief review of the related gauge theory and D-brane bound states, we explain the construction of the crystal model as the main method of the asymptotic analysis of the instanton sector.
\subsection{D-branes bound states, gauge theory on divisors and instantons}
We consider D4-D2-D0 branes bound states on toric Calabi-Yau threefolds $X$, consisting of a single D4-brane wrapped on a non-compact toric divisor of the Calabi-Yau, and arbitrary number of D2-branes wrapped on the compact two-cycles and D0-branes as point particles in the Calabi-Yau.
By dimensionally reducing $X$, the 4d, $\cN=2$ supersymmetric $U(1)$ gauge theory is obtained and the D4-D2-D0 bound states can be seen as the BPS states of the gauge theory, carrying the charge $\gamma=(n, \beta)$ with $n$ and $\beta$ being the number of D0 and D2 branes. The spectrum of the BPS bound states is captured by the so-called divisor \textit{BPS index} $\Omega_\cD(\gamma)$.

The low-energy effective gauge theory on the D4-brane wrapping the toric divisor is the 4d topologically twisted $\cN=4$ super Yang-Mills theory, called \textit{Vafa-Witten theory}. The D2-D0 BPS bound states in D4-brane are the solutions of the (anti)self-dual Yang-Mills equation, and they are called instantons of the Vafa-Witten theory. The instanton counting in this gauge theory on the 4d part of the D4 brane in the Calabi-Yau can be performed via the enumeration of the D4-D0 BPS states.
The above theory is an example of the geometric construction of the instantons in type II string theory, where a configuration of $N$ $Dp$-branes with $n$ $D(p-4)$-branes is considered. The low energy theory in the world-volume of $N$ $Dp$-branes is the $(p+1)$-dimensional $\cN=4$, $U(N)$ super Yang-Mills theory and the instanton in the background $Dp$-brane is the $D(p-4)$-brane. The gauge theory on $D(p-4)$-brane has $U(n)$ gauge symmetry while the $U(N)$ symmetry on the  $Dp$-branes are considered as the flavour symmetry. These configurations describe the Open strings stretching between $D(p-4)$- and $Dp$-branes.

In order to count the D-branes BPS bound states we need a generating function, called the BPS generating function,
\bea
\label{GF}
\cZ_\cD(q,Q)=\sum_{n,\beta}\Omega_\cD(n, \beta)\ q^{n}Q^{\beta},
\eea
where the $q$ and $Q$ are the fugacity factors associated with the D0 and D2 branes respectively, and $Q^\beta=\prod_jQ_j^{\beta_j}$, with the product over the two-cycles of the Calabi-Yau.

On the other hand, the BPS generating function is the instanton partition function, and the BPS index is given by the topological Euler characteristic of the moduli space of instantons $\cM_{n,\beta}^{\text{inst}}(\cD)$,
\bea
\Omega_\cD(n, \beta)=\chi(\cM_{n,\beta}^{\text{inst}}(\cD))= \int_{
\cM_{n,\beta}^{\text{inst}}(\cD)} \text{eul}\left(T \cM_{n,\beta}^{\text{inst}}(\cD)\right),
\eea
where $\text{eul}\left(T \cM_{n,\beta}^{\text{inst}}(\cD)\right)$ is the Euler class of stable tangent bundle, for a review see for example section (4.1) in \cite{ci-sz}.

Our main goal in this article is to count the BPS states/instantons and compute the BPS index/ Euler characteristic. In particular we are interested in the asymptotic behavior of the BPS index in the large $n$ limit, and possible phase transitions in the system. The first step towards this goal is to develop the asymptotic counting tools such as crystal models and their generating functions which are appropriate for the study of the $\cN=4$ toric quiver quantum mechanics, as the low energy effective theory on the D-branes. 
In the following, we summarize the construction of the toric quiver and associated two-dimensional crystal melting model, following \cite{Nish}.
\subsection{Quiver, dimer model and two-dimensional crystal model on divisors}
\label{sec:quiv crys}
This part is a brief review of essential constructions of the quiver, dimer model and the two-dimensional crystal models associated with the D4-D2-D0 bound states and related gauge theories, introduced in \cite{Nish}.

The quiver is a directed graph denoted by $Q=(Q_0, Q_1, Q_2)$ as a set of nodes $Q_0$, arrows $Q_1$ and faces $Q_2$ in the graph. The nodes of the quivers are associated with the gauge groups $U(N_i)$ at each node $i$. The field content of the gauge theory is also encoded in the quiver, and the arrows between the nodes of the quiver are the chiral multiplets $X_a\in Q_1$ in the bifundamental representation. Moreover, the possible interactions of the gauge theory is explained by the superpotential of the gauge theory and it is given by the sum over the trace of the product of the chiral multiplets around the faces of the quiver.

The dual graph of the quiver is called the brane tiling or the dimer model. The brane tiling is a combinatorial method to construct and represent the D-branes configurations and associated gauge theories. The brane tiling $Q'$ is a bipartite graph and the duality implies $Q_0'=Q_2$, $Q_1'=Q_1$, and $Q_2'=Q_0$. In this work we mostly consider the dimer models with isoradial embedding. This is an embedding of the dimer model on the torus such that every vertex of the dimer model at a boundary of a face is on a unit circle \cite{Ken-iso}. The tropical geometry associated with the statistical mechanics of the dimer model plays the key role in the asymptotic analysis of the quiver.

The quiver quantum mechanics in the low energy limit of the D4-D2-D0 branes is obtained from the quiver of the D6-D2-D0 branes by replacing the flavour D6 brane with a flavour D4-brane wrapped on a toric divisor $\cD$ of the Calabi-Yau and applying the associated constraints. Let us briefly review the constraints from the embedding of the D4 flavour brane. Suppose there are two D2-brane nodes $i$ and $j$ with a chiral fields $X_F$ associated with the arrow from $i$ to $j$, and the D4-brane node $*$ is adjacent to $i$ and $j$. The brane tiling implies that there are two massless quark and antiquark $I$ and $J$ attached to D4-brane, associated with the arrows between $*$ and $i$, and $j$ and $*$. Therefore, in addition to the superpotential $W_0$ made out of the product of chiral multiplet $X_a$, the presence of the flavour D4-brane in brane tiling, leads to an extra term $W_f$, to the superpotential
\bea
W=W_0+W_f,\quad W_f= J X_F I.
\eea
The F-term relations from the superpotential of the quiver are
\bea
\frac{\partial W_0}{\partial X_a} &=&0\quad \text{for} \quad X_a\ne X_F, \quad \frac{\partial W_0}{\partial X_F} + IJ =0,\nonumber\\
\frac{\partial W_f}{\partial I} &=&J X_F=0, \quad \text{and}\quad \frac{\partial W_f}{\partial J} =X_FI=0.
\eea

As we review in the following, the dimer model plays a crucial role in the identification of the constraints on the instanton moduli space. A bipartite dimer model is a set of black and white nodes connected with the oriented edges. The perfect matching $m$ in the dimer model $Q'$ is the set of all oriented edges such that every vertex in $Q_0'$ is covered by one and only one oriented edge in $m$. 
The perfect matchings are in bijection with the lattice points of the Newton polygon $\Delta$ and the corners of $\Delta$ are in one-to-one correspondence with the toric divisors, thus there is unique perfect matching $m_\cD$ for any divisor $\cD$.
At low energy limit, the BPS index becomes the Witten index of the quiver. To compute the Witten index, we need to study the moduli space of the vacua $\cM_{\text{D4}}$ of quiver quantum mechanics on D4-D2-D0 via the quiver representation. The moduli space $\cM_{D4}$, as a subspace of $\cM_{D6}$ of original D6-D2-D0, is invariant under the torus action of a $U(1)$-subgroup of $U(1)^2\times U(1)_R\simeq U(1)^3$, depending on the divisor $\cD$. For further explanations on the symmetries of the problem consult with chapter 3 in \cite{Nish}.
The F-term constraints imposed on the moduli space $\cM_{D4}$ are studied in \cite{Nish}, and we summarize them without proof in the following,
\bea 
J=0\quad \text{which leads to} \quad \frac{\partial W_0}{\partial X_F} =0,\quad \text{and} \quad X_a=0\ \ \text{if}\ \ \psi(X_a)\in m_\cD,
\eea
where $\psi:Q\to Q'$ is the map from quiver to its dual, the dimer model.
To study $\cM_{D4}$ and compute the Witten index, we need to elaborate on the path algebra associated with the quiver $\mathbb{C}Q$. The elements of this algebra are the paths between nodes of the quiver and the product of the elements is the concatenation of the paths. The factor algebra is obtained by imposing the F-term relations on the path algebra, $A=\mathbb{C}Q/F$, where $F$ is the ideal generated by all the F-terms.
The factor algebra associated with the divisor quiver is the original D6-D2-D0 factor algebra of the quiver but further constrained by the F-term conditions implied by the superpotential of the flavour D4 brane.
The $\cM_{D4}$ is identified with the factor algebra imposed by D-term relations, and it is sometimes called stable A-module.

The two-dimensional crystal model is constructed on the universal cover of the dimer or equivalently its dual, the quiver $\tilde Q$, by using the path algebra. Any path from a reference node $i_0\in \tilde Q_0$ to a arbitrary node $j\in \tilde Q_0$ is given in the form $v_{i_0 j} \omega^l$, where $v_{i_0 j}$ is the shortest path between $i_0$ and $j$, and $\omega$ is a loop around a face of the quiver. However, for the paths in the A-module $\cM_{D4}$, it is shown in \cite{Nish}, that $l=0$, and all the paths crossing $m_\cD$ are eliminated. The F-equivalent classes of all the paths which do not cross $m_\cD$ form a set, and the $U(1)^2$-fixed points of $\cM_{D4}$ are in one-to-one correspondence with the finite ideals of this set. The original crystal is constructed by stacking atoms on the end points of the paths $v_{i_0 j} \omega^l$, on the nodes $j$ of the quiver, and at the depths $l$ inside the 3d crystal model, and thus all the elements of the A-module $\cM_{D_4}$ will lie on the 2d plane of depth zero, i.e. one of the facets of the 3d crystal model. Therefore the two-dimensional crystal model is constructed on $\tilde{Q}$ by the elements
of the factor algebra which are the shortest paths $v_{i_0 j}$, via putting an atom at the ending point of the path.

A particular facet of 3d crystal model is identified with a chosen divisor and the precise shape of the 2d crystal model is determined by some oriented paths in the dimer model, called zig-zag paths. The zig-zag paths are the paths that turn maximally left at white vertices and maximally right at black vertices. In fact, they are the external legs of the toric diagram and the ridges of the 3d crystal model. Thus, any facet of the crystal is bounded by two zig-zag paths. 

Finally, we have all the elements to discuss the BPS generating function. At low energy, the BPS index becomes the Witten index of the quiver quantum mechanics and it can be computed by the sum over the fixed-points of the torus-action $U(1)^2$ on $\cM_{D4}$. These fixed points are the ideals of the A-module and the molten configurations $\lambda$ of the 2d crystal model. Thus, the generating function of BPS states on divisor \eqref{GF} is given by the partition function of the 2d crystal model,
\bea
\cZ_\cD(q, Q)=\cZ_{\text{CM}}\left(q_1, q_2, ...\right)=\sum_{ \lambda}\prod_{k\in Q_0} q_k^{|\lambda_k|},
\eea
where the sum is over the molten configurations $\lambda$ with $|\lambda_i|$ being the number of atoms in the configuration $\lambda$ at node $i$, i.e. the rank of the gauge group associated with the node $i$ at the configuration $\lambda$, and $q_i$ is fugacity factor associated to node $i$.

Having reviewed the construction of the crystal model, we can move towards the next step and the heart of this study which is the asymptotic analysis of the BPS states/instantons on the divisors. Based on the counting tools of the crystal models and by introducing asymptotic methods from tropical geometry, large deviation technique and number theory, we propose our new approach towards the thermodynamics and phase structure of the instanton sector, in the following section.
\section{Asymptotic analysis and phase structure of instantons on divisors}
In this part, we explain the main results of this paper about the asymptotic analysis of the quiver and the phase structure of the instanton sector.
The first issue in the asymptotic analysis of the crystal model is about the definition of the asymptotic limit and the appropriate scaling. The coupling $g$ of the crystal model, defined as $g=-\log q$, can be seen as the lattice spacing of the crystal. The number of the boxes in the crystal is given by $n$, thus in order to take the continuum and thermodynamics limits at the same time, we need to $g\to 0$ and $n \to \infty$ in an appropriate way. A plausible approach to do this is by fixing the volume of the crystal $g^2 n$ at some large value. Using the dimer model parameters, the original toric quiver is on the $M\times M$ cover of the torus. In the thermodynamic limit, the quiver is enlarged on the universal cover, $M\to \infty$ such that $g M$ is fixed but large. This is consistent with the scaling $M^2\sim n$.
\subsection{Amoeba and asymptotics}
The main object in the asymptotic analysis of the two-dimensional crystal melting model and the associated quiver is the limit shape. The limit shape is a smooth convex function which is obtained, in the asymptotic limit, by rescaling the profile function of the crystal, denoted by $\psi(x)$. This is the profile of the ranks of the gauge groups of the quiver. Let us define the Amoeba of the Newton polynomial $P(z,w)\in \mathbb{C}[z^\pm, w^\pm]$, as
\bea
\cA(P)= \left\{\left(\log |z|, \log |w|\right) | P(z,w)=0\right\}.
\eea
Our first observation is that the limit shape of the 2d crystal model on each facet of the 3d crystal model is the boundary of the Amoeba of the Newton polynomial of the dimer model associated with that facet, and it is denoted by $\cA^B$. This can be argued by the fact that the limit shape of the 3d crystal model is given by the Ronkin function and, roughly speaking, the Amoeba is the projection of the Ronkin function in the plane \cite{Ke-Ok-Sh}. Thus, it is easy to see that the limit shape of the 2d crystals on the facets of the 3d crystals are the boundaries of the Amoeba.
For the practical and computational purposes, we can choose any of the boundaries of the Amoeba, however, it is more convenient to choose, if possible, the one which after some reflections and/or rotations resembles the limit shape of the random partitions in the up-right quarter of the plane. This will become clear once we consider explicit examples in chapter \ref{Examples}.

In the asymptotic limit, the volume factor $q^n$ in generating function \eqref{GF} can be written as an integral of the limit shape, using the scaling relation $M^2 \int_{0}^1 \cA^B(x) \ dx = n$. Similarly, the degeneracy factor in the generating function can be formally written as an integral of the entropy function denoted by $\sigma(s)$. This is the tension of the 2d crystal, a function of the slope of profile of the quiver. We will elaborate on the entropy in the next part. But before getting into the details, notice that in the asymptotic limit, the generating function of the crystal model, by the WKB approximation, is given by a continuous integral in terms of the dominant configuration, the limit shape and its tension, as
\bea
\label{BPS gf}
\cZ_\cD (M) \sim \exp{\left(M \int_0^1 \sigma_\cD (s)\ dx - g M^2 \int_0
^1\cA^B(x)\ dx\right)}.
\eea
As we observe in the above result, the partition function of the 2d crystal model in the asymptotic limit is dominated by the limit shape contribution and thus it is plausible to make an analogy with the asymptotics of the three-dimensional crystal model \cite{Ya-Oo}.

\subsection{Entropy density and Legendre transform}
The slope function of the profile $\psi(x)$ of the quiver is defined as
\bea
\label{slope 2d}
\frac{\partial \psi(x)}{\partial x}= s(x).
\eea
The tension of the quiver is a function of slope and is obtained from the Legendre transform of the profile
\bea
\label{entropy 2d}
\sigma_\cD (s) =\cL[\psi]=\sup_{\psi}\left[ \psi(x) - x\ s(x) \right]= \cA^B(x) - x\ s^*(x),
\eea
where the slope function is given by $s^*(x)=\partial \cA^B(x)/\partial x$. Furthermore, Legendre duality implies $\partial \sigma_\cD(s)/\partial s= x(s)$. For more convenience we drop the $*$ in the slope function, from now on. As we expect, the limit shape of the crystal or the boundaries of the Amoeba are the maximizer of the entropy.
The idea is that the tension of the limit shape of the crystal, or in other words, the profile of the quiver is the entropy of the BPS states.
Moreover, the instanton growth rate, defined as the logarithm of the degeneracy, is given by the number of fluctuating profiles lying close to the limit shape. As we explained briefly, the asymptotics of the instanton growth rate can be obtained from the asymptotics of the partition function. In fact, the first integral in generating function \eqref{BPS gf}, is an area term, thus its second root is the correct dimension for the number of fluctuating profiles and thus using the scaling relation $n\sim M^2$, the instanton growth rate becomes
\bea
\label{BPS GR 2d}
\log\Omega_\cD(n) \sim n^{\frac{1}{2}}\left(\int \sigma_\cD(s) \ dx\right)^{\frac{1}{2}},
\eea
where the integral is evaluated along a part of the spine of the Amoeba or the toric diagram that is captured by the limit shape which is associated with the chosen divisor $\cD$. In order to compute the above integral and obtain the numerical factor, one needs to compute the entropy as a function of coordinate $x$ and evaluate the integral. This can be done by first finding the entropy as a function of slope, which is explained in the following, and then transforming the entropy to a function of coordinate by using the explicit slope function in terms of coordinate.
Indeed, as we will see in section \ref{Saddle-Point Analysis in Reduced Quiver}, the saddle-point analysis of the quiver, leads to a parallel result up to a factor two, and both results actually lead to the numerical factor for the instantons growth rate.
Notice that, since the number of D2 branes is fixed in the asymptotic limit, the $Q^l$ term has finite fixed contribution to  the generating function \eqref{GF} and does not contribute in the asymptotic formula, except than a constant.

As we explained, in the asymptotic limit, the tension of the crystal model is given by the Legendre transform of the limit shape. Moreover, from the properties of the Legendre transform, we have
\bea
-\frac{\partial \sigma_\cD}{\partial s} \circ s= \text{Id},
\eea
which can be solved to explicitly compute the entropy
\bea
\label{ent from inverse}
\sigma_\cD(s)=- \int x(s)\ d s,
\eea
where $x(s)$ is the inverse function $s^{-1}(x)$. Thus, we find that the Legendre transform of the Amoeba of the quiver gives the explicit functional form of the entropy as in Eqs. \eqref{entropy 2d} and \eqref{ent from inverse}. We will obtain explicit formulas in some examples in chapter \ref{Examples}.

\subsection{Free energy and saddle-point analysis}
The free energy of the quiver is obtained via the integral of the limit shape of the 2d crystal model, the boundary of the Amoeba. There are two different ways to compute the area which is captured by the boundary of the amoeba, either by direct computations of the integral of the Amoeba boundaries or via the tropical geometry and toric geometry.
To compute the free energy, $\cF= \log \cZ$, we start with Eq. \eqref{BPS gf},
\bea
\cF_\cD (M)\sim \left(M \int_0^1 \sigma_\cD (s)\ dx - g M^2 \int_0^1\cA^B(x)\ dx\right).
\eea
Then by using the Legendre duality between tension and the limit shape,
\bea
\cA^B(g M x)= \sigma_\cD (s) + g M x\ s(x),\quad \dfrac{d (x\cA^B(x))}{dx}= \cA^B(x)+x\ s(x),
\eea
and the fact that the total derivative term in the integral can be put to zero as a boundary term, we have
\bea
\cF_\cD (M)\sim M\int_0^1 (\sigma_\cD (s) - g M \cA^B(x))\ dx= M\int_0^1(\sigma_\cD (s) + g M x s)\ dx= M\int_0^1\cA^B(g M x)\ dx,
\eea
and finally by a change of variable we have
\bea
\cF_\cD(g) \sim \frac{1}{g} \int \cA^B (x) \ dx.
\eea
In the above equation, the area under the limit shape is the area trapped between the boundary of the Amoeba and the spines of the Amoeba surrounding a divisor. 
This result is consistent with the scaling relation $M\sim g^{-1}$, obtained earlier.

The actual computation of the area of the limit shape and free energy depends on the geometry of the Newton polygon and is a practical matter, However, as we mentioned earlier, in some symmetric cases we can compute the free energy of a divisor more explicitly using the toric geometry results. Let $\Delta$ be the Newton polygon of a Newton polynomial $P(z,w)$ defined as a convex hull of the exponents of the monomials with nonzero coefficients. Then results in \cite{Mi-Ru} and \cite{Ke-Ok-Sh} imply that the area enclosed by the Amoeba of the dimer, as a Harnack curve, is given by the area of the corresponding Newton polygon $\Delta$,
\bea
\label{Am ineq}
\text{Area}(\cA) = \pi^2 \text{Area}(\Delta).
\eea
Further studies about the Amoeba and its analytic aspects can be found in \cite{pa-ru,yge}. Further relations between the dimer mode and Harnack curves is studied in \cite{ke-ok2} and it is found that the isoradial dimer models are characterized by the genus zero Harnack curves.

In the case of symmetric toric diagrams and Amoebas, the above result leads to
\bea
\cZ_\cD(g)\sim \exp{(\frac{1}{g}\int \cA^B(x) \ dx)}=\exp{(\frac{1}{g}A(\cA^B))}= \exp{(\frac{\pi^2}{l g}A(\Delta))},
\eea
where we denote the area by $A$ and we used the symmetry of the Amoeba to observe that the area enclosed by the Amoeba is $l$ times bigger than the area enclosed between each boundary of the Amoeba and the coordinates, with $l$ being the number of the tentacles of the Amoeba. We will see some examples of symmetric Amoebas in chapter \ref{Examples}.

The total free energy, as the sum of the free energies associated with all the lattice points on the exterior of the Newton polygon (including corner points which are associated with the divisors of the Calabi-Yau and non-corner points) can be computed as the sum of the captured areas by all the boundaries of the Amoeba, 
\bea
\label{total F}
\cF^{(t)}(g)= \sum_{i} \cF^{(i)}(g)\sim\sum_i \frac{1}{g}\int \cA^{B_i}(x) \ dx= \frac{1}{g} A(\cA)= \frac{\pi^2}{g} A(\Delta),
\eea
where $i$ runs over all the lattice points of the exterior of the Newton polygon or equivalently all the boundaries of the Amoeba.
The free energy on each divisor might depends on the K\"{a}hler parameters $Q_i$'s but the total free energy \eqref{total F} is independent of the K\"{a}hler moduli $Q_i$'s of the Calabi-Yau and in fact is a topological invariant under any deformations of the Calabi-Yau and their K\"{a}hler parameters.
As the total free energy contains contributions from non-divisor points, the physical interpretation of this observation is not clear to the author. However, in the case of non non-divisor points in Newton polygon, i.e. all the lattice points are located in a corner of the Newton polygon, the total free energy has the instanton total free energy interpretation.

In the tropical limit, $Q_i\to 0$, the Amoeba decomposes to the isolated $\mathbb{C}^3$ vertices. However, the area of the Amoeba is a topological invariant and independent of the $Q_i$'s, and thus by using the fact that the area of the Amoeba of each vertex is $A(\cA_v)=\pi^2 A(\Delta_v)=\pi^2/2$, the area of Amoeba of any Calabi-Yau can be written as $A(\cA)=\sum_{v} A(\cA_v)= \frac{\pi^2}{2}c_v$, where $c_v$ denotes the number of vertices in the toric diagram.

\subsection*{Saddle-point analysis and instanton growth rate}
\label{Saddle-Point Analysis in Reduced Quiver}
In this part, we apply the saddle-point analysis to compute the explicit numerical factor of the instanton degeneracy and growth rate. Using the saddle-point analysis and the limit shape method, the instanton growth rate on a divisor becomes,
\bea
\label{Cauchy}
\Omega_\cD(n)= \frac{1}{2\pi i} \oint dq\ \cZ_\cD \ q^{-n-1}=\frac{1}{2\pi i} \oint dq\ \exp\left(\frac{1}{g}A(\cA^B)\right) \ q^{-n-1} =\frac{1}{2\pi i} \oint dq\ e^{h(g)},
\eea
where $h(g)= ng+ \frac{1}{g} A(\cA^B)$.
In the symmetric case, by using the toric geometry, the asymptotic degeneracy can be obtained by $h(g)= n g+\frac{\pi^2}{l g}A(\Delta)$. By extremization of $h(g)$, we obtain the extremizing value of the coupling,
$g^*= \pi\sqrt{\frac{A(\Delta)}{l n}}$,
at which we have
\bea
h'(g^*)=0, \quad h(g^*)=2\pi \sqrt{\frac{n A(\Delta)}{l}} ,\quad h''(g^*)=\frac{2}{\pi}\sqrt{\frac{l n^3}{A(\Delta)}}.
\eea
After a Gaussian integration in Eq. \eqref{Cauchy}, the instanton degeneracy can be obtained as
\bea
\label{deg asymp}
\Omega_\cD(n)\sim \frac{e^{h(g^*)}}{\sqrt{2\pi h''(g^*)}}=\frac{1}{2}(\frac{A(\Delta)}{l n^3})^{1/4}\exp{(2\pi \sqrt{n A(\Delta)/l})},
\eea
and thus instanton growth rate, formally related to the entropy as in Eq. \eqref{BPS GR 2d}, can be computed in the symmetric cases from the area of the Newton polygon as 
\bea
\label{inst deg}
\log\Omega_\cD(n)\sim n^{\frac{1}{2}}\left(\int \sigma_\cD(s) \ dx\right)^{\frac{1}{2}}\sim n^{\frac{1}{2}} \left(4\pi^2 A(\Delta)/l\right)^{\frac{1}{2}}\sim 2\sqrt{A(\cA^B)}\ n^{\frac{1}{2}}.
\eea
The above result for the instanton growth rate, as stated in terms of the area of the boundary of the Amoeba, $\log\Omega_\cD(n)\sim 2\sqrt{A(\cA^B)}\ n^{\frac{1}{2}}$ is in general valid for symmetric cases as well as non-symmetric cases.
The above result can also be obtained using the fact that the entropy as the Legendre dual of the free energy can be extremised and given by the free energy which is the area of the limit shape. 

Similar to the total free energy discussed before, we can think of the total Instanton growth rate on all the divisors, as a topological invariant of the Calabi-Yau which is independent of the K\"{a}hler parameters. By using the saddle-point result we can compute the total growth rate as
\bea
\log \Omega^{(t)} (n)\sim n^{\frac{1}{2}} \left(4\pi^2 A(\Delta)\right)^{\frac{1}{2}}.
\eea
For more convenience, we drop the divisor subscript $\cD$ in the rest of this paper, unless it is necessary.

\subsection{Phase structure and Gumbel distribution}
In this part, we adopt and generalize some results in number theory to explain the phase structure related to the Hagedorn phase transition and instanton condensation. This phase structure is basically originated from the asymmetric Gumbel distribution of fluctuations in the 2d crystal model, as we will explain in the following.

The first heuristic observation is the convergence/divergence phase transition in the partition function at a critical point. Following the scaling relations explained in the beginning of this chapter, by setting $n\sim g^{-2}$ with the proportionality factor as $\beta$, and using the instanton growth rate in Eq. \eqref{inst deg}, we observe that the partition function \eqref{GF}, modulo the finite factor $Q^\beta$, behaves in the asymptotic limit as
\bea
\label{Hagedorn}
\cZ(\beta)\sim\sum_n \exp\left[-(\beta-\beta_c)\sqrt{n}\right], \quad \beta_c=2 \sqrt{ A(\cA^B)},
\eea
where the critical inverse temperature of the Hagedorn-like phase transition is obtained from the asymptotic behavior of the degeneracy, Eq. \eqref{deg asymp}. We'll denote the critical area which is the area of the limit shape by $A_c$, and the critical inverse temperature by $\beta_c=2A_c^{\frac{1}{2}}$. Roughly speaking, the partition function is convergent in the low temperature regime $\beta>\beta_c$ and it diverges at high temperature $\beta<\beta_c$.

In order to study the phase structure in a precise manner, we need to study the fluctuations around the limit shape that causes the phase transition. To regularize the fluctuations, we consider the restricted crystal model. In other words, we put the system in a box of length $N$ and arbitrary width. In other words, there is an upper bound on the length of the first row of the crystal model. Then, we study the largest part of the system which is the first row of the crystal model and its fluctuations in the thermodynamic limit $N\to \infty$. This leads to the study of the finite size effects of the restricted model in contrast to the unrestricted model and exploration of the associated phase transition.
\subsubsection*{Gumbel fluctuation distribution}
The precise asymptotic analysis of the restricted partitions with uniform measure, performed in \cite{er-le}, provides a mathematical framework for the study of the phase structure in the 2d crystal models. A straightforward generalization of the Erd\"{o}s-Lehner result for the integer partitions, as random partitions with a uniform measure, to the general 2d crystal model can be formulated as follows. As $n\to \infty$, (the upper bound on) the length of the first row of the crystal tends to infinity, $N\sim \frac{1}{c}\sqrt{n}\log n$, and more precisely, based on the analogy, we conjecture that the degeneracy of the crystal model, which is the BPS index, in the finite box behaves as
\bea
\lim_{n\to \infty}\frac{\Omega_N(n)}{\Omega(n)}= e^{-c^{-1}e^{-cx}},
\eea
where $\Omega(n)$ is the degeneracy of the unrestricted crystal, $x= \frac{N}{\sqrt{n}}- \frac{1}{2c}\log n$ and the constant $c$ is the square root of the area of the limit shape and depends on the geometry of the crystal, $c=\pi\sqrt{A(\cA^B)}$. In the original setting for the integer partition, the result is proved for the constant $c= \pi/\sqrt{6}$. One naturally expects such a generalization, as the 2d crystal model is a geometric generalization of the partitions and the statistical weight is a uniform measure, the volume of the crystal.
To rephrase this result in the canonical ensemble, let us consider the crystal model with uniform measure $P_n(\lambda)= \frac{1}{\Omega(n)}$ in the canonical ensemble. Then, 
The longest row of the crystal model has the Gumbel distribution \cite{er-le},
\bea
\lim_{n\rightarrow \infty}\textit{Prob}\ \left(\frac{\lambda_1}{\sqrt{n}}- \frac{1}{2c}\log n\leq x\right)= e^{-c^{-1}e^{-cx}}.
\eea

As we saw, one of the physical implications of this result for the crystal model is that the size of the box tends to infinity as $N\sim \sqrt{n}\log n$ and it would be natural to fix the finite ratio $\frac{N}{\sqrt{n}\log n}\equiv |\gamma|$. In fact, the re-scaled first row $\frac{\lambda_1}{\sqrt{n}}$ does not have a finite average in the large $n$ limit and it diverges logarithmically, $\langle\frac{\lambda_1}{\sqrt{n}}\rangle= \frac{1}{2c}\log n$. This is unlike the random partitions with Plancherel measure with $\langle\frac{\lambda_1}{\sqrt{n}}\rangle= 2$, but this is consistent with the fact that the tails of the limit shape do not cross the $x,y$ axis at finite number. However, the area of the limit shape is a finite parameter in the crystal model and it can be proposed as the phase transition order parameter, as we will see in the following.

In the grand canonical ensemble with volume weight $q^{|\lambda|}$, for the restricted partition function we have
\bea
\cZ_{N}(q)=\textit{Prob}\ \left(\lambda_1\leq N\right)=\sum_{\lambda, \lambda_1\leq N} q^{|\lambda|}= \sum_n\sum_{\lambda \vdash n, \lambda_1\leq N} \Omega_N(n) q^n.
\eea
One can reformulate the results of \cite{er-le} in the grand canonical ensemble as explained in \cite{be-bo},
\bea
\lim_{q\rightarrow 1^-}\textit{Prob}\ \left(\lambda_1 |\log q| - \frac{|\log q|\log (1-q)}{\log q}\leq x\right)= e^{-e^{-x}},
\eea
or equivalently, in terms of the partition function
with $q=e^{-g}$, we have
\bea
\lim_{g\to 0}\frac{\cZ_N(g)}{\cZ(g)}= e^{-e^{-x}},
\eea
where $\cZ(g)$ is the normalization factor, i.e. the partition function where the restriction on the first row is relaxed and $x= N g + \log g$.
This result is consistent with the results in the canonical ensemble, using the scaling relation $n= c^2 g^{-2}$.

\subsubsection*{Phase structure}
The asymptotic distribution of the fluctuations in the above results is called Gumbel distribution. This is an asymmetric distribution. As we will see, this asymmetry is the origin of the phase transition in the system.
In the grand-canonical ensemble, for the free energy defined by $\cF_N=-\lim_{n \to \infty}\log \cZ_{N}(g)$, we have 
\bea
\lim_{g\to 0}\ (\cF_N(g)- \cF(g))= e^{-x}, \quad x= N g + \log g.
\eea
Fixing the ratio $\frac{Ng}{\log g}= -2c|\gamma|$, we observe
\bea
\lim_{g\to 0}\ (\cF_N(g)- \cF(g))= \lim_{g\to 0} \ g^{(2c|\gamma|-1)}=
\begin{cases}
\infty & \text{for }\quad |\gamma|< \frac{1}{2c}\\
0 & \text{for }\quad |\gamma|> \frac{1}{2c}
\end{cases}.
\eea
Thus, we observe that the there is a critical point $|\gamma^*|=\frac{1}{2c} = \frac{1}{2}A_c^{-1/2}$, at which the fluctuation contribution to the free energy jumps from zero to infinity. In other words, the finite size effects, the infinite contribution is caused by the fluctuations around the limit shape for $A>A_c$, otherwise ($A<A_c$) there is no finite size effect and the free energy is the free energy of the limit shape configuration $\cF(g)$. 

Similarly, in the canonical ensemble for the entropy $S_N=\log \Omega_{N}(n)$ we have
\bea
\lim_{n\to \infty}\ (S_N(n)- S(n))= \frac{1}{c}e^{-cx}, \quad x= \frac{N}{\sqrt{n}} - \frac{1}{2c} \log n,
\eea
and for the fixed ratio $\frac{N}{\sqrt{n}\log n}=|\gamma|= \frac{1}{2} A^{-1/2}$, we obtain
\bea
\lim_{n\to \infty}\ (S_N(n)- S(n))= \lim_{n\to \infty} \ -\frac{1}{c}\ n^{\frac{1}{2}(1-(\frac{A}{A_c})^{-1/2})}=
\begin{cases}
\infty & \text{for }\quad A>A_c\\
0 & \text{for }\quad A<A_c
\end{cases}.
\eea
There exists similar interpretation for the fluctuation contribution to the entropy around the limit shape and the system acquires a tension associated with the limit shape. Furthermore, notice that in the regime with the non-zeros fluctuation contribution to the entropy, naturally it is subleading to the surface tension of the limit shape $S(n)\sim n^{1/2}$, as we have 
\bea
\lim_{n \to \infty}\frac{S_N(n)}{S(n)}\sim n^{-\frac{1}{2}(\frac{A}{A_c})^{-1/2}}\to 0.
\eea
It is important to mention that the above results for the limit shape fluctuation contributions are valid in the vicinity of the limit shape, just before and after its formation.

Having explained the mathematical observation about the singular points in the asymptotic limit, at which the entropy and free energy diverge, in the next part we discuss and elaborate on some possible interpretations of these results.
\subsubsection*{Hagedorn phase transition and instanton condensation}
Possible interpretation of the above results concerns the instantons and its moduli space in the asymptotic regime. The phase transition happens at the large number of instantons and thus it is associated with the behavior of the gas of instantons on the divisors of the Calabi-Yau threefold. There are some possible physical phenomena happening in the gas of instantons such as condensation of the instantons and formation of the instanton condensate. As we explain in the following, similar condensation of the instantons associated with the formation of the limit shape in the crystal model, is possibly responsible for the Hagedorn phase transition. However, as it is shown in \cite{ver-BE}, there is no Bose-Einstein condensation in the 2d random partitions and instantons.

As we explained in the beginning of this section, there is a Hagedorn phase transition at some critical temperature $\beta_c^{-1}$ between the strong and weak coupling regimes. We observed that the critical temperature is closely related to the limit shape area and the phase transition is associated with the formation of the limit shape and the fluctuations around that. In fact, the Hagedorn phase transition is consistent with the idea that the limit shape formation and developing a "surface" tension in the crystal is associated with a phase transition and moreover the limit shape itself is the critical one- and two-dimensional hypersurface separating the frozen and smooth phases in the 2d and 3d crystal models, respectively.

To interpret the Hagedorn phase transition in the instanton sector caused by the Gumbel fluctuations, the key point is to focus on the density of instantons. The meaning of the Hagedorn phase transition in the instanton sector is that in the high temperature phase $\beta<\beta_c$, the number of BPS states grows with energy and we have Hagedorn density of BPS states. In fact, for the instanton density defined as $\rho= n/N^2$, by using the Erd\"{o}s-Lehner scaling $N\sim ( A(\cA^B))^{-1/2}\sqrt{n}\log n$, we observe that $\rho\sim A \log^{-2} n$ and thus, after re-scaling, one can define a regularized density $\tilde\rho$ as the area of the profile function and the critical density as the area of the limit shape $\tilde\rho_c= A_c$. Thus, the limit shape formation and the fluctuations about it can be seen as the condensation of the instantons and fluctuations around the condensate in the high density phase $\tilde\rho >\tilde\rho_c$.

The Gumbel distribution implies that after the instanton condensate forms, there is fluctuation around the instanton condensate. This originates from the existence of the finite size effect of the condensate and leading to the finite size corrections to the free energy and entropy in this phase. Notice that the phase transition extracted from the Gumbel distribution happens exactly at the same critical temperature of the Hagedorn phase transition. This is natural since we speculated that the Hagedorn phase transition emerged from the fluctuation patterns around the limit shape.

Bearing in mind the relation between integer partitions and bosonic strings on the one hand, and the relation between the instantons, D0-D4 states and bosonic strings on the other hand, the above phase transition seems to be closely related to the Hagedorn phase transition in the bosonic string theory \cite{at-wi}. It is also possibly connected to the similar phase structure in the gas of D-branes and D-instantons in \cite{va1,va2,gre}, and similar asymptotic results for the Betti numbers and Euler characterisitcs in \cite{hau}.

From another perspective, as we mentioned before, similar to the 3d crystal melting model, the 2d crystal model has a smooth phase inside the Amoeba and frozen phases in the unbounded complement components of the Amoeba. The smooth phase is the fluctuating phase with infinite number of instantons whereas the frozen phase has zero number of instantons.

Having introduced a new approach towards the asymptotic analysis of the quiver gauge theory on the divisors of the Calabi-Yau threefolds, in the following section we implement our methods in some concrete examples and obtain explicit results. 
\section{Examples}
\label{Examples}
In this chapter We study some examples of quiver with the Amoebas and Harnack curves of genus zero and one, such as $\mathbb{C}^3$, conifold, local $\mathbb{P}^1\times \mathbb{P}^1$ and local $\mathbb{P}^2$ quivers, and study their asymptotics and thermodynamics.
In each example we compute the free energy, entropy function, and growth rate, using the Amoeba and its properties.
Where possible, we compare the free energy, entropy and growth rate obtained from our geometric and analytic methods with the results obtained from standard asymptotic analysis of the explicit generating functions. 
\subsection{$\mathbb{C}^3$ divisors}
\begin{figure}
\centering
\includegraphics{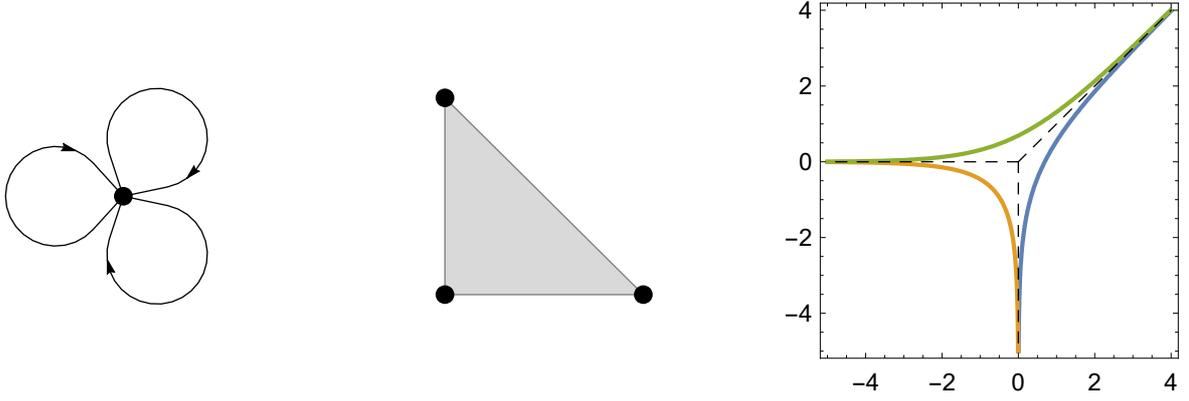}
\caption{Left: $\mathbb{C}^3$ quiver, middle: Newton polygon, right: Amoeba and its spines}
\label{C3amoeba}
\end{figure}

We start with the fundamental example of $\mathbb{C}^3$ divisor. This example is the most studied example as it is the integer partition or the 2d random partition with uniform measure. The quiver, Newton polygon and Amoeba with spines of this example is illustrated in Fig. \ref{C3amoeba}.
The Newton polynomial of $\mathbb{C}^3$ is $P(z,w)= -1+z+w$, and the equations of the boundaries the Amoeba in Fig. \ref{C3amoeba}, are $\color{brown} e^x+e^y=1$, $\color{OliveGreen} -e^x-e^y=1$, $\color{NavyBlue} e^x-e^y=1$. We choose the yellow boundary of the amoeba and after the reflections $x\to -x$ and $y\to -y$, it becomes $-1+e^{-x}+e^{-y}=0$. Thus,
the limit shape, as the boundary of the Amoeba, can be obtained by solving the equation for $y(x)$ and then denoting $y$ by $\cA^B(x)$,
\bea
\label{c2 LS}
\cA^B_{\mathbb{C}^3}(x)= -\log(1-e^{-x}).
\eea
This matches with the well-known result for the limit shape of the uniform random partitions $\exp(-\frac{\pi x}{\sqrt{6}})+ \exp(-\frac{\pi y}{\sqrt{6}})=1$ in which the coordinates are normalized by a numerical factor (square root of area) so that the limit shape has the unit area.
The free energy of $\mathbb{C}^2$ quiver which lives on a divisor of $\mathbb{C}^3$, can be computed from the area of the limit shape and Newton polygon,
\bea
\cF_{\mathbb{C}^3}(g)=\frac{1}{g} \int \cA^B_{\mathbb{C}^2}(x) dx = \frac{\pi^2}{3g}A(\Delta_{\mathbb{C}^3})=\frac{\pi^2}{6g}.
\eea
This result can be obtained directly by evaluating the area under the limit shape,
\bea
\cF_{\mathbb{C}^3}(g)&=& \frac{1}{g}\int_0^\infty -\log(1-e^{-x})\ dx\nonumber\\
&=& \left.-\frac{1}{g}\left(\frac{x^2}{2}+ x\log(1-e^{-x})-x\log(1-e^x)-\text{Li}_2(e^x)\right)\right\vert_{0}^{\infty}= \frac{\pi^2}{6g}.
\eea
Using other plausible method, we obtain
\bea
\cF_{\mathbb{C}^3}(g)= \frac{1}{g}\int_0^\infty -\log(1-e^{-x})\ dx= \frac{1}{g}\sum_{k=1}^\infty \int_0^\infty \frac{e^{-kx}}{k} \ dx = \frac{1}{g}\sum_{k=1}^\infty \frac{1}{k^2} = \frac{\pi^2}{6g}.
\eea
Consider the integer partition generating function,
\bea
\cZ (q) = \sum_{n=0}^{\infty} p(n) q^n = \prod_{i=1}^\infty \frac{1}{1-q^i}.
\eea
Then our result for the free energy matches, up to leading order, with the following result from the probability theory of the uniform partitions, for a summary consult with \cite{zho},
\bea
\label{C2 Free}
\log \cZ (q) = -\frac{c^2}{\log q} + \frac{1}{2}\log \frac{-\log q}{2\pi} + O(|\log q|),
\eea
where $c=\sqrt{\zeta(2)}=\pi/\sqrt{6}$.

In the following we obtain the entropy of $\mathbb{C}^2 $ crystals from the Legendre transform of their limit shapes. 
The entropy from the Legendre dual of the limit shape $\cA_B^{\mathbb{C}^3}(x)$ is given by
\bea
\label{C2 entropy}
\sigma_{\mathbb{C}^3}(s)=\cA^B_{\mathbb{C}^3}(x) - x\ s(x).
\eea
We can use the limit shape equation \eqref{c2 LS}, to obtain 
\bea
\label{c2 slope}
s(x)=\frac{-e^{-x}}{1-e^{-x}},\quad
x(s)= \log(\frac{s-1}{s}), 
\eea
and thus write the entropy in terms of the slope. In a parallel approach, the entropy can be obtained by integrating $\frac{\partial \sigma}{\partial s}= x$, implied from the Legendre duality,
\bea
\label{C2 shannon}
\sigma_{\mathbb{C}^3}(s)=-\int x(s) ds= s \log s+ \log(1- s)-s\log(s-1).
\eea
It is easy to check that the results from computing $\sigma(x)$ in right hand side of Eq. (\ref{C2 entropy}) and in Eq. (\ref{C2 shannon}) match. Using Eqs. \eqref{C2 shannon} and \eqref{c2 slope} in Eq. \eqref{BPS GR 2d}, we obtain the instanton growth rate
\bea
\label{c2 deg from entropy}
\log \Omega_{\mathbb{C}^3}(n)\sim n^{\frac{1}{2}} \left(\int_0^\infty\sigma_{\mathbb{C}^3}(s) \ dx\right)^{\frac{1}{2}}=\pi \sqrt{\frac{n}{3}}.
\eea
On the other hand, the BPS growth rate from the saddle-point analysis is
\bea
\log \Omega_{\mathbb{C}^3}(n)\sim n^{\frac{1}{2}} \left(4\pi^2 A(\Delta_{\mathbb{C}^3})/l_{\mathbb{C}^3}\right)^{\frac{1}{2}}.
\eea
Evaluating $A(\Delta_{\mathbb{C}^3})/l_{\mathbb{C}^3}= 1/6$, we observe that
saddle-point result matches with the Hardy-Ramanujan asymptotics of the integer partition,
\bea
p(n)=\frac{1}{4\pi\sqrt{3}}e^{\pi\sqrt{\frac{2n}{3}}}(1+O(n^{-\frac{1}{2}})).
\eea
The direct computation of the growth rate from the entropy function in Eq. \eqref{c2 deg from entropy} also matches with the above result, modulo a factor $\sqrt{2}$. We will see in the following that the same numerical factor appears in other examples, and that is suggestive of the universality of this factor. 

Total free energy and growth rate of the $\mathbb{C}^3$ divisors, are
\bea
\cF^{(t)}_{\mathbb{C}^3}(g)\sim \frac{\pi^2}{g}A(\Delta_{\mathbb{C}^3})=\frac{\pi^2}{2g},\quad \log \Omega^{(t)}_{\mathbb{C}^3}(n)\sim n^{\frac{1}{2}} \left(4\pi^2 A(\Delta_{\mathbb{C}^3})\right)^{\frac{1}{2}}=\pi \sqrt{n}.
\eea
\begin{figure}
\centering
\includegraphics{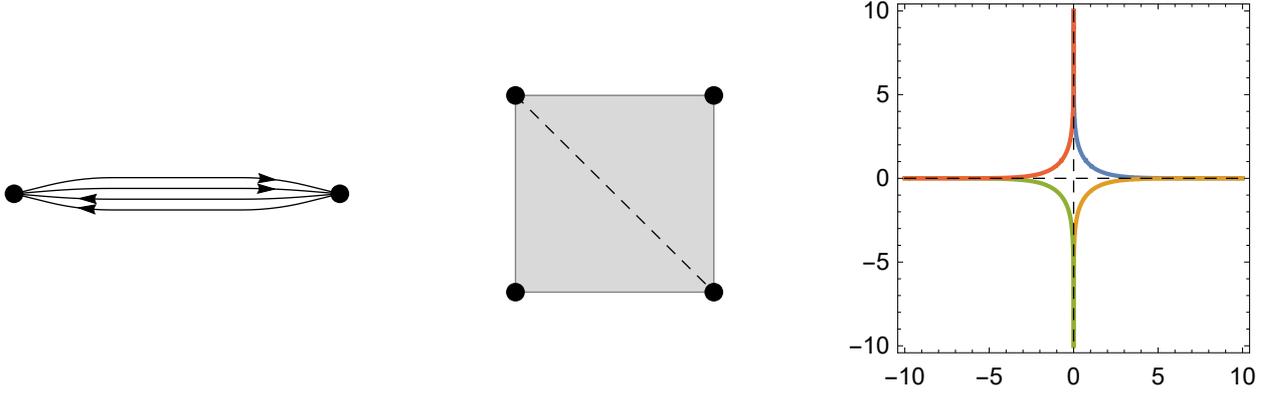}
\caption{Conifold quiver, Newton polygon and the Amoeba with the spines at $Q=1$}
\label{conif}
\end{figure}

Finally, the inverse critical temperature, Eq. \eqref{Hagedorn}, in $\mathbb{C}^3$ quiver is $\beta_c=2\pi/ \sqrt{ 6}$.

\subsection{Conifold divisors}
The second example is the conifold divisor. The crystal model associated with the conifold quiver is the pyramid partition and its dimer model is the Aztec diamond model \cite{you}. We study the statistical mechanics of the 2d crystal model that lives on the facets of the pyramid partition. The Newton polynomial of the conifold quiver is $P(z,w)= -1+z+w+Qzw$, where $Q$ is the K\"{a}hler parameter, which is related to the geometry of the toric diagram by $Q=e^{-t}$ with $t$ being the length of the internal leg of the conifold. The conifold quiver, its Newton polygon and the Amoeba with the spines are shown in Fig. \ref{conif}. The Newton polynomial determines the equation of the boundaries of the Amoeba and we choose the blue boundary of the Amoeba in Fig. \ref{conif}, with the following equation to study, $-1+e^{-x}+e^{-y}+ Q e^{-x} e^{-y} =0$. Then, from this equation, the boundary of the Amoeba which is the limit shape of the associated facet of the pyramid partition can be obtained as
\bea
\label{C LS}
\cA^B_{\cC}(x;Q)=-\log\Big(\frac{1-e^{-x}}{1+Q e^{-x}}\Big).
\eea

The free energy can be obtained directly by evaluating the area under the limit shape of the resolved conifold,
\bea
\cF_{\cC}(g;Q)&=& \frac{1}{g}\int_0^\infty -\log\Big(\frac{1-e^{-x}}{1+Q e^{-x}}\Big)\ dx\nonumber\\
&=&\left. \frac{1}{g} \left(x\log(1-e^x)-x\log\left(\frac{e^x-1}{e^x+Q}\right)-x\log\left(\frac{e^x+Q}{Q}\right)\right)\right\vert_0^\infty\nonumber\\
&+&\left. \frac{1}{g}\left(\text{Li}_2(e^x)-\text{Li}_2\left(-\frac{e^x}{Q}\right) \right)\right\vert_0^\infty\nonumber\\
&=&\frac{1}{g}\left(\pi^2/6- \text{Li}_2(-Q)\right).
\eea
Other plausible method produces the same result
\bea
\label{C free en}
\cF_{\cC}(g;Q)&=& \frac{1}{g}\int_0^\infty -\log\Big(\frac{1-e^{-x}}{1+Q e^{-x}}\Big)\ dx= \frac{1}{g}\sum_{k=1}^\infty \int_0^\infty \left(\frac{e^{-kx}}{k}-\frac{e^{-kx}(-Q)^k}{k}\right) \ dx\nonumber\\
&=& \frac{1}{g}\sum_{k=1}^\infty \left(\frac{1}{k^2}-\frac{(-Q)^k}{k^2}\right) =\frac{1}{g}\left(\pi^2/6- \text{Li}_2(-Q)\right).
\eea
\begin{figure}
\centering
\includegraphics{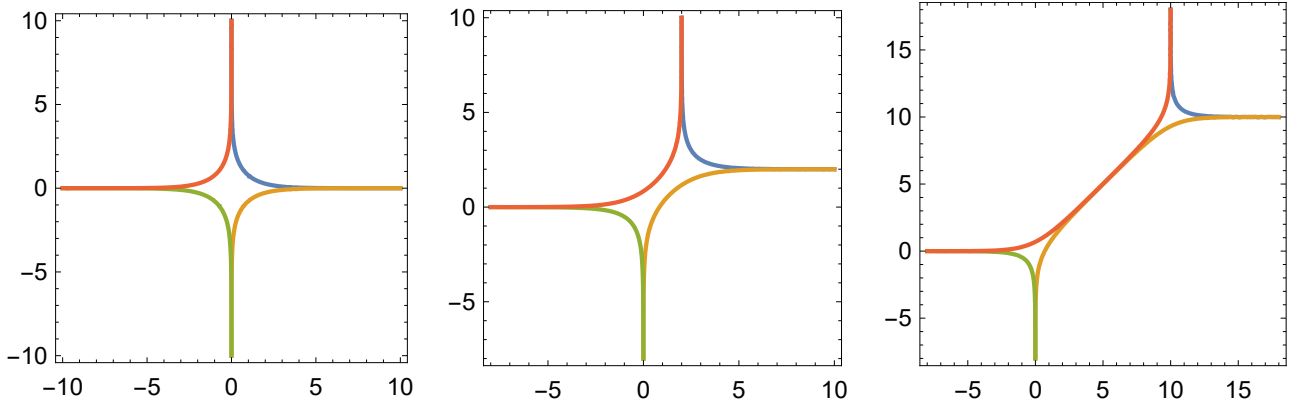}
\caption{$\cC$ Amoeba at different values of $Q$, left: $Q=1$, middle: $Q>1$, right: $Q\gg 1$}
\label{conifamoeba}
\end{figure}
At $Q=1$, the free energy $\cF_{\cC}(g;Q=1)$, as obtained above, reproduces the free energy obtained from the area of the Amoeba,
\bea
\label{conif F}
\cF_{\cC}(g;1)=\frac{1}{g} \int\cA^B_{\cC}(x;Q=1)\ dx = \frac{\pi^2}{4g}A(\Delta_{\cC})=\frac{\pi^2}{4g}.
\eea
Alternatively, we can compute the free energy by using the standard asymptotic analysis of the generating function for the conifold divisor obtained in \cite{Nish}, as
\bea
\cZ_\cC(q,-Q)= \prod_{i=1}^\infty \left(\frac{1}{1-q^i} \right) \prod_{j=0}^\infty (1-Qq^j),
\eea
and using the leading order in Eq. \eqref{C2 Free}, or by replacing the sum with the integral and direct computation, the free energy defined by $\mathcal{F}_\cC(q,-Q)
=\log\cZ_\cC(q,-Q)$, becomes
\bea
\mathcal{F}_\cC(q,-Q)
&=& -\frac{\pi^2}{6\log q}+ \log \prod_{j\geq 0}(1-Q q^j)= -\frac{\pi^2}{6\log q} +\sum_{j\geq 0}\log (1-Q q^j)\nonumber\\
&=& -\frac{\pi^2}{6\log q} + \sum_{j\geq 0}\sum_{k \geq 1}-\frac{Q^k q^{jk}}{k}= -\frac{\pi^2}{6\log q} -\sum_{k \geq 1}\frac{Q^k}{k}\int_0^\infty e^{-jk g} dj\nonumber\\
&=& -\frac{\pi^2}{6\log q} -\sum_{k \geq 1}\frac{Q^k}{k^2 g}= -\frac{\pi^2}{6\log q} + \frac{\text{Li}_2(Q)}{\log q}.
\eea
This is consistent with the earlier result in Eq. \eqref{C free en}. 

\begin{figure}
\centering
\includegraphics[width=10cm]{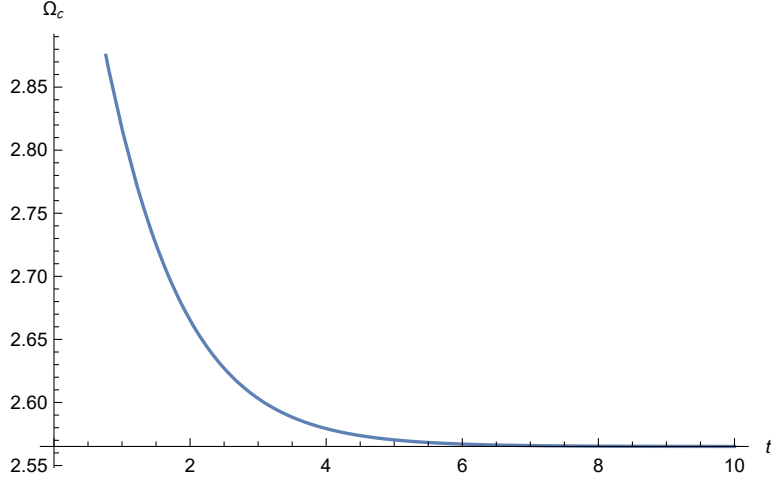}
\caption{Instanton growth rate of the conifold} 
\label{conifgrowth}
\end{figure}

In the following, we obtain the entropy of the conifold divisors from the Legendre dual of their limit shapes.
Using the limit shape we can obtain the slope function and its inverse,
\bea
s(x)&=& \Big(\frac{1}{1-e^x}-\frac{Q}{e^x+Q}\Big),\nonumber\\
x(s)&=& \log(\frac{-1 + s - Q - s Q + 
\sqrt{4 Q s^2 + (-1-Q+s-Qs^2)^2}}{2 s}).
\eea
The entropy function is obtained from the inverse function of the slope as,
\bea
\sigma_\cC(s;Q)&=&-\int x(s)\ ds\nonumber\\
&=&\log\Big(1-s-Q(1+s)- \sqrt{(1+Q)((-1+s)^2+Q(1+s)^2)}\Big)\nonumber\\
&-& 
s \log\Big(\frac{-1 + s + Q(1+s) + \sqrt{(1+Q)((-1 + s)^2 + (1 + s)^2 Q) }}{2 s}\Big).
\eea
The growth rate of the conifold divisor can, in principle, be computed via the entropy function, however it is easier to obtain the explicit results by using the free energy in Eq. \eqref{C free en} in the saddle-point method,
\bea
\log \Omega_\cC(n;t)\sim 2\left(\frac{\pi^2}{6}- \text{Li}_2(-e^{-t})\right)^{\frac{1}{2}}n^{\frac{1}{2}}.
\eea
The numerical factor of the growth rate as the function of $t$ is depicted in Fig.\ref{conifgrowth}. Notice that a $t\to\infty$ one can reproduce the growth rate of $\mathbb{C}^3$ divisor.

At $Q=1$, the slope of the limit shape and its inverse become
\bea
s(x)=\frac{-2}{-e^{-x}+e^x},\quad x(s)= \log\left(\frac{-1-\sqrt{1+s^2}}{s}\right).
\eea
The entropy function is obtained as
\bea
\sigma_\cC(s) = -\int x(s) ds = -s \log\left(-\frac{1+\sqrt{1+s^2}}{s}\right)- \log(s+\sqrt{1+s^2}),
\eea
and thus the instanton growth rate is 
\bea
\log \Omega_{\cC}(n)\sim n^{\frac{1}{2}} \left(\int_0^\infty\sigma_{\cC}(s) \ dx\right)^{\frac{1}{2}}=\pi \sqrt{\frac{n}{2}}.
\eea
On the other hand, the growth rate from the saddle-point analysis is obtained as
\bea
\log \Omega_{\cC}(n)\sim n^{\frac{1}{2}} \left(4\pi^2 A(\Delta_{\cC})/l_{\cC}\right)^{\frac{1}{2}}=\pi \sqrt{n},
\eea
and as we expect, the growth rates obtained from two methods agree up to a numerical factor $\sqrt{2}$.

The Amoeba of the conifold depends on the K\"{a}hler parameter as depicted in Fig. \ref{conifamoeba}, but the area of the amoeba remains constant by changing $Q$. Thus, the total free energy and growth rate of the conifold, for any $Q$, are
\bea
\cF^{(t)}_{\cC}(g)\sim \frac{\pi^2}{g}A(\Delta_{\cC})=\frac{\pi^2}{g},\quad \log \Omega^{(t)}_{\cC}(n)\sim n^{\frac{1}{2}} \left(4\pi^2 A(\Delta_{\cC})\right)^{\frac{1}{2}}=2\pi \sqrt{n}.
\eea
We can consider different divisors of the conifold at any $Q$. In any divisor of the conifold, using the limit shape of that divisor we can compute the free energy, entropy and growth rate. Alternatively, by using the total free energy of the conifold given by the area of the Amoeba and the free energy of the other divisors, we can simply compute them. At $Q>1$, let's call the divisor that we considered so far $\cD_1$, this is the divisor associated with the green or blue boundaries of the Amoeba in Fig. \ref{conifamoeba}, and then study another divisor called $\cD_2$, which is associated with the red or orange boundaries of the Amoeba. In this divisor, taking into account the symmetry of the toric diagram, the free energy and growth rate can be obtained from the free energy of the $\cD_1$ divisor and total free energy,
\bea
\cF_{\cD_2}(g;Q)= 1/2\ \cF^{(t)} - \cF_{\cD_1}(g;Q)= \frac{\pi^2}{2g} - \cF_{\cD_1}(g;Q)= \frac{1}{g}\left(\pi^2/3+ \text{Li}_2(-Q)\right).
\eea
\bea
\log \Omega_{\cD_2}(n;t)\sim 2\left(\frac{\pi^2}{3}+ \text{Li}_2(-e^{-t})\right)^{\frac{1}{2}}n^{\frac{1}{2}}.
\eea

Finally, the inverse critical temperature of the resolved conifold quiver divisors can be computed from Eq. \eqref{Hagedorn} as
\bea
\beta_c^{\cD_1}(t)= 2 \left(\pi^2/6- \text{Li}_2(-e^{-t})\right)^{\frac{1}{2}},\quad \beta_c^{\cD_2}(t)= 2 \left(\pi^2/3+ \text{Li}_2(-e^{-t})\right)^{\frac{1}{2}},
\eea
and at $Q=1$, on both divisors we have $\beta_c=\pi$.

\subsection{Local $\mathbb{P}^1\times \mathbb{P}^1$ divisors}
\begin{figure}
\centering
\includegraphics{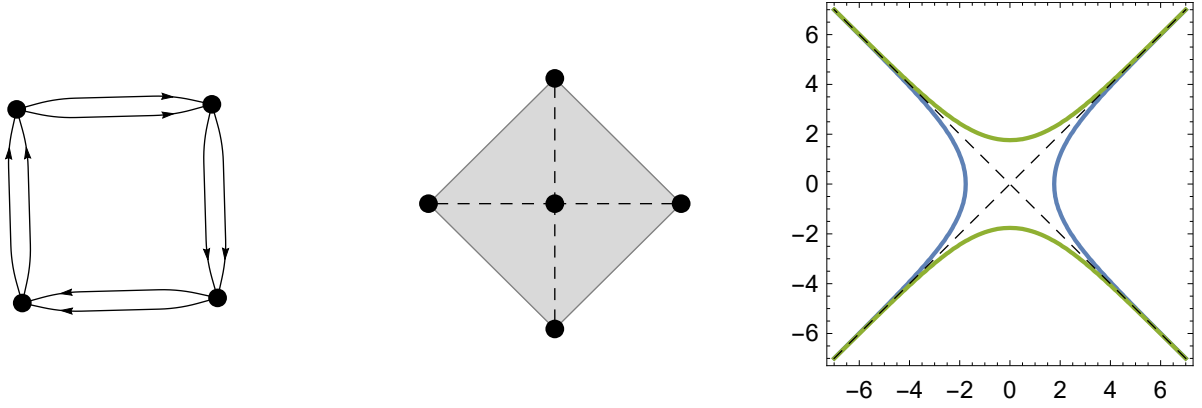}
\caption{Local $\mathbb{P}^1\times \mathbb{P}^1$ quiver, Newton polygon and the Amoeba with spines at $k=4$}
\label{F0}
\end{figure}

\begin{figure}
\centering
\includegraphics{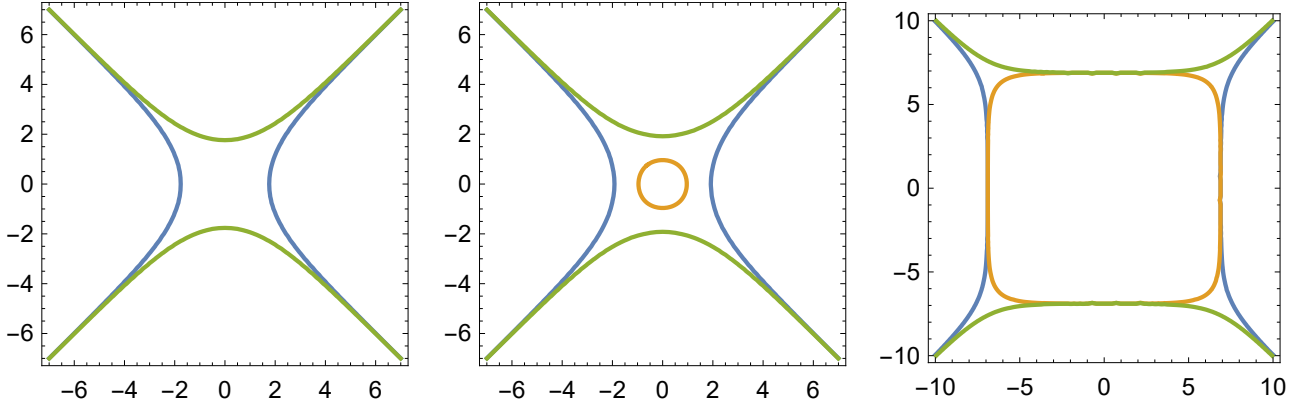}
\caption{Local $\mathbb{P}^1\times \mathbb{P}^1$ Amoeba at different $k$, left: $k=4$, middle: $k>4$, right: $k\gg 4$}
\label{F0amoeba}
\end{figure}

Having discussed two examples with the known generating functions, the third example is the local $\mathbb{P}^1\times \mathbb{P}^1$, known as Hirzebruch quiver $\mathbb{F}_0$, and its generating function is unknown. Thus, we can only apply the geometric approach in the asymptotic analysis of this quiver.
The Newton polynomial of this quiver is $P(z,w) = -k+ z+w+1/z+1/w$. The quiver, Newton polygon and its Amoeba at the isoradial point $k=4$, and the spines are depicted in Fig. \ref{F0}. The size of the bounded component (hole) of the Amoeba increases with $k$, for $k>4$, see Fig. \ref{F0amoeba}.

The limit shape associated with the right blue boundary of the Amoeba, in Fig. \ref{F0}, is computed via solving the spectral curve $e^y+e^{-y} -e^x-e^{-x} -k =0$, and after a rotation by $\pi/4$, to fit into the $(x,y)$ coordinate system, it is
\bea
\cA^B_{\mathbb{F}_0}(x;k)=\sqrt{2}\log\left(\frac{k e^{x/\sqrt{2}}+\left(4-8e^{\sqrt{2} x}+ 4e^{2\sqrt{2}x}+k^2 e^{2\sqrt{2}x}\right)^{1/2}}{2(-1+e^{\sqrt{2}x})}\right).
\eea
In the following, we focus on the isoradial limit of this quiver.
At the isoradial point, $k=4$, the limit shape is
\bea
\cA^B_{\mathbb{F}_0}(x;4)=\sqrt{2}\log\left(\frac{1+ e^{x/\sqrt{2}}}{-1+e^{x/\sqrt{2}}}\right).
\eea
The free energy at $k=4$, can be obtained by computing the area under the limit shape,
\bea
\cF_{\mathbb{F}_0}(g;4)&=& \frac{1}{g}\int_0^\infty \sqrt{2}\log\left(\frac{1+ e^{x/\sqrt{2}}}{-1+e^{x/\sqrt{2}}}\right)\ dx\nonumber\\
&=& \left.\frac{1}{g} \left(-2\sqrt{2}\ x \Arctanh[e^{x/\sqrt{2}}] +\sqrt{2}\ x \log\left(\coth(\frac{x}{2 \sqrt{2}})\right)\right)\right\vert_{0}^{\infty}\nonumber\\
\ &+&\left.\frac{1}{g} \left( 4\ \text{Li}_2( e^{x/\sqrt{2}}) - \text{Li}_2(e^{\sqrt{2} x})\right)\right\vert_{0}^{\infty}\nonumber\\
&=& \frac{\pi^2}{2g}.
\eea
Using another plausible approximation method explained in the conifold quiver, one can compute the above integral and reproduce the same result. Moreover,
the above result is also consistent with the free energy computed from the area of the Amoeba, using the symmetry of the Amoeba at any $k$,
\bea
\cF_{\mathbb{F}_0}(g;k)=\frac{1}{g} \int\cA^B_{\mathbb{F}_0}(x) dx = \frac{\pi^2}{4g}A(\Delta_{\mathbb{F}_0})=\frac{\pi^2}{2g}.
\eea
and thus, as the area of the Amoeba only depends on the area of the Newton polygon and not the K\"{a}hler parameter $k$, we observe that the free energy is a deformation invariant quantity, and we have $\cF_{\mathbb{F}_0}(g;k)=\cF_{\mathbb{F}_0}(g;4)$.

From the limit shape we can compute the slope function and its inverse,
\bea
s(x)=\frac{-2}{e^{-x/\sqrt{2}}+e^{x/\sqrt{2}}},\quad x(s)=\sqrt{2} \log\left(\frac{-1-\sqrt{1+s^2}}{s}\right).
\eea
Thus, the entropy function is obtained from the integral of the inverse of the slope function,
\bea
\sigma_{\mathbb{F}_0}(s;4) =- \int x(s) ds = -\sqrt{2}\ s \log\left(-\frac{1+\sqrt{1+s^2}}{s}\right)-\sqrt{2} \log(s+\sqrt{1+s^2}).
\eea
The growth rate is computed via the entropy function
\bea
\log \Omega_{\mathbb{F}_0}(n;4)\sim n^{\frac{1}{2}} \left(\int_0^\infty\sigma_{\mathbb{F}_0}(s;4) \ dx\right)^{\frac{1}{2}}=\pi \sqrt{n}.
\eea
In consistency with the computation of the growth rate in the saddle-point analysis, we have
\bea
\log \Omega_{\mathbb{F}_0}(n;4)\sim n^{\frac{1}{2}} \left(4\pi^2 A(\Delta_{\mathbb{F}_0})/l_{\mathbb{F}_0}\right)^{\frac{1}{2}}=\pi \sqrt{2n}.
\eea

Amoeba of the quiver has fixed tentacles independent of $k$, however, as we mentioned, the bounded component of the Amoeba which is a hole inside the Amoeba emerges for $k>4$ and it grows as a function of $k$, see Fig. \ref{F0amoeba}. Although the size of the hole changes with $k$, but the area of the Amoeba remains constant, $A(\cA_{\mathbb{P}^1\times \mathbb{P}^1})= \pi^2 A(\Delta_{\mathbb{P}^1\times \mathbb{P}^1})=2\pi^2$. Finally, the total free energy and growth rate on all four divisors of the $\mathbb{F}_0$ quiver, for any $k$, can be obtained as
\bea
\cF^{(t)}_{\mathbb{F}_0}(g)\sim \frac{\pi^2}{g}A(\Delta_{\mathbb{F}_0})=\frac{2\pi^2}{g},\quad \log \Omega^{(t)}_{\mathbb{F}_0}(n)\sim n^{\frac{1}{2}} \left(4\pi^2 A(\Delta_{\mathbb{F}_0})\right)^{\frac{1}{2}}=2\pi \sqrt{2n}.
\eea
The inverse critical temperature of $\mathbb{P}^1\times \mathbb{P}^1$ quiver is given by Eq. \eqref{Hagedorn} as $\beta_c=2\pi/ \sqrt{ 2}$ at any $k$.

\subsection{Local $\mathbb{P}^2$ divisors}
\begin{figure}
\centering
\includegraphics{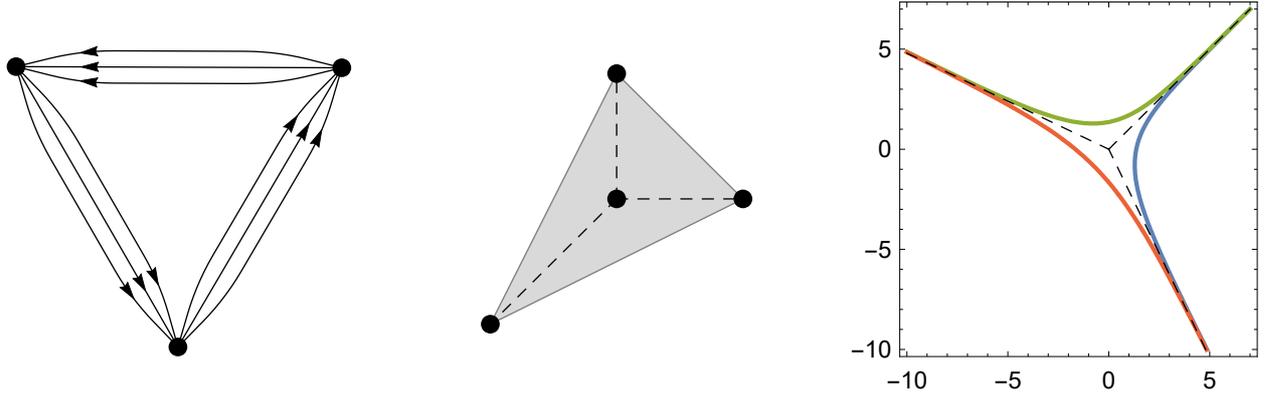}
\caption{$\mathbb{P}^2$ quiver, the Newton polygon and the Amoeba and its spines at $Q=1/27$}
\label{P2}
\end{figure}

In this part we consider the second example of quivers with a Newton polygon which has an inside point, called local $\mathbb{P}^2$ quiver. The Newton polynomial of this quiver is $P(z,w)= 1+z+w+ \frac{Q}{zw}$ and the dual graph of the Newton polygon, the toric diagram and its Amoeba are illustrated in Fig. \ref{P2}. There are two phases of the quiver, the phase $Q>1/27$, in which the Amoeba has no hole, corresponding to a genus zero Harnack curve and the phase  $Q<1/27$, in which there is a hole inside the Amoeba and corresponding to a genus one Harnack curve. 

\begin{figure}
\centering
\includegraphics{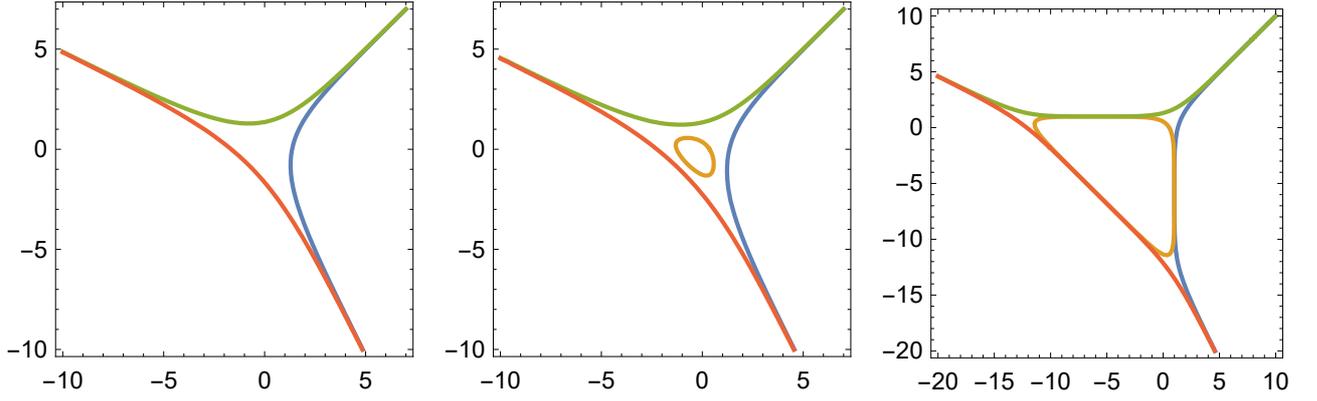}
\caption{$\mathbb{P}^2$ Amoeba different K\"{a}hler parameters: left: $Q=1/27$, middle: $0<Q<1/27$, right: $0<Q\ll 1/27$ }
\label{P2amoeba}
\end{figure}

We consider the blue boundary of the Amoeba in Fig. \ref{P2amoeba}, as the limit shape for a generic $Q$ and we obtain, from the solutions of the spectral curve $e^x-e^y-Q\ e^{-x-y}=1$,
\bea
\cA_{\mathbb{P}^2}^B(x)=\log\left(\frac{1}{2}e^{-x}\left(-e^x(1-e^x)\pm e^{x/2}\sqrt{-4Q+e^x-2e^{2x}+e^{3x}}\right)\right),
\eea
where plus/minus sign gives the two complementary parts of the limit shape, meaning that it is made of union of these two separate curves. In this example, and by using the above spectral curve, the center of the Amoeba which is the point that the hole emerges is located at $(-1,-1)$. However, we expect that the center and the degenerate point of the Amoeba is located at the origin $(0,0)$, as illustrated in Fig. \eqref{P2}. In order to put the center of the Amoeba at the origin, we shift $x\to x-1$ and $y\to y-1$. However, this shift does not change neither the area of the Amoeba (limit shape) nor the slope function and the entropy. 

Taking into account the symmetry of the Amoeba, the free energy, independent of $Q$, can be evaluated from the area of the Amoeba, and thus for any $Q$ we have
\bea
\cF_{\mathbb{P}^2}(g)=\frac{1}{g} \int_{}\cA^B_{\mathbb{P}^2}(x)\ dx = \frac{\pi^2}{3g}A(\Delta_{\mathbb{P}^2})=\frac{\pi^2}{2g}.
\eea
The free energy can be also obtained directly by computing the area under the limit shape, however for the general $Q$ the computation is involved.
The slope function is computed from the limit shape as
\bea
s(x;Q)=-\frac{1}{2}\pm \frac{e^{x/2}(3 e^x-1)}{2\sqrt{-4Q+ e^x(e^x-1)^2}},
\eea
where $\pm$ refers to two separate parts of the limit shape, as explained before.
In principle, the inverse function of the slope of both parts of the limit shape can be computed explicitly, although computationally involved. However, at critical K\"{a}hler parameter, $Q_c=1/27$, the computations can be simplified and the inverse function of the slope for both parts of the limit shape is obtained as
\bea
x(s)= \log\left(\frac{(1+2s)^2}{3(s^2+s-2)}\right).
\eea
The entropy function can be computed from the slope as,
\bea
\sigma_{\mathbb{P}^2}(s;Q_c)=-\int x(s) ds= -\log\left(\frac{(1-s)(1+2s)}{(2+s)^2}\right)-s \log\left(\frac{(1+2s)^2}{3(s^2+s-2)}\right). 
\eea
Using the entropy function, in principle one can compute the instanton growth rate and similar to previous examples, we expect to have
\bea
\label{P2growth}
\log \Omega_{\mathbb{P}^2}(n)\sim n^{\frac{1}{2}} \left(\int\sigma_{\mathbb{P}^2}(s) \ dx\right)^{\frac{1}{2}}=\pi \sqrt{n},
\eea
however, the actual computation of the integral to find the numerical value is tedious and instead we can use an alternative method.
More explicitly, using the saddle-point analysis we can compute the growth rate, for any $Q$, as
\bea
\label{P2saddle}
\log \Omega_{\mathbb{P}^2}(n)\sim n^{\frac{1}{2}} \left(4\pi^2 A(\Delta_{\mathbb{P}^2})/l_{\mathbb{P}^2}\right)^{\frac{1}{2}}=\pi \sqrt{2n},
\eea
which we have the agreement between Eqs. \eqref{P2growth} and \eqref{P2saddle}, up to a constant numerical factor $\sqrt{2}$.

Finally, we compute the total free energy and growth rate on all the divisors. The tentacles of the Amoeba is fixed and independent of K\"{a}hler parameter, but the size of the hole inside the Amoeba increases as $Q$ tends to zero. In Fig. \ref{P2amoeba}, we demonstrate the Amoeba with three different values for the K\"{a}hler parameter. However the area of the Amoeba is independent of $Q$ and thus for the total free energy and growth rate of the local $\mathbb{P}^2$ quiver we have,
\bea
\cF^{(t)}_{\mathbb{P}^2}(g)\sim \frac{\pi^2}{g}A(\Delta_{\mathbb{P}^2})=\frac{3\pi^2}{2g},\quad \log \Omega^{(t)}_{\mathbb{P}^2}(n)\sim n^{\frac{1}{2}} \left(4\pi^2 A(\Delta_{\mathbb{P}^2})\right)^{\frac{1}{2}}=\pi \sqrt{6n}.
\eea
The inverse critical temperature of $\mathbb{P}^2$ quiver is obtained from Eq. \eqref{Hagedorn} as $\beta_c=2\pi/ \sqrt{2}$, for any $Q$.

\section{Conclusion and discussion}
In this study, the asymptotic analysis of the quiver gauge theories associated with the divisors of the Calabi-Yau is performed by using methods and results from tropical geometry, large deviation techniques and number theory. Consequently, the explicit results for the free energy, entropy and growth rate of instantons are obtained. We observed that the total free energy and the total growth rate as the sum of contributions from all the divisors of the Calabi-Yau threefold, is proportional to the area of the Amoeba. Therefore, as explicitly stated in the examples, the larger area of the Newton polygon is, the larger free energy and entropy we have. More explicitly, in terms of the total instanton growth rate and inverse critical temperature we observe
\bea
\Omega^{(t)}_{\mathbb{P}^1\times \mathbb{P}^1}>\Omega^{(t)}_{\mathbb{P}^2}>\Omega^{(t)}_{\cC}>\Omega^{(t)}_{\mathbb{C}^3}, \quad 
\beta_c^{\mathbb{P}^1\times \mathbb{P}^1}=\beta_c^{\mathbb{P}^2}>\beta_c^{\cC}>\beta_c^{\mathbb{C}^3}.
\eea

The original 3d crystal model is interpreted as the discrete building blocks of the Calabi-Yau manifolds since the limit shape of the crystal model, given by the Ronkin function, is the smooth mirror geometry of the Calabi-Yau threefold, \cite{Ya-Oo}. In the same spirit, we have the limit shape of the 2d crystal model given by the Amoeba, which is the solution of the mirror curves of the Calabi-Yau. Thus, the 2d crystal model can be seen as the discretization of the smooth geometry of the toric divisors of the Calabi-Yau.

The extension of our methods to study the asymptotics of the orbifold quivers, as an important infinite class of quivers, is highly interesting, from a physical and mathematical point of view. There are extensive studies on the D-brane bound states on the $\mathbb{C}^2/\mathbb{Z}_N$ orbifold and the instantons on the resolved $A_{N-1}$ ALE spaces \cite{Do-Mo}, and also possible gravity duals and black holes \cite{He-Va}. We can directly apply our method in this class of quivers and study their thermodynamics and interpret the results for the black holes.

The main focus of this study is the isoradial quivers. The non-isoradial quivers are an interesting class of quivers and their asymptotics are described by the Amoebas with bounded components, and Harnack curves of the genus higher than zero. From the physical point of view, this class of quivers contain the gas phase inside the holes of the Amoeba which has contributions to the entropy density. Computation of the contributions of the gas phase to the entropy is an interesting and challenging question.

There are related studies on the thermodynamics and phase structure of the $SU(N)$ Vafa-Witten theory on K3 surfaces in the large $N$ limit and BTZ black holes \cite{pap,oku}. It would be interesting to generalize the construction in \cite{Nish} to include $N$ D4-branes wrapped on the toric divisors and then adopt the similar asymptotic methods to study the phase structure of these theories.

Since the BPS generating function in most of the quivers is not known, using the geometric approach developed in this article, one can study the asymptotic aspects of these quivers, and the obtained result would be useful in the study of the generating functions towards their finding. 

An interesting direction for the future studies would be to generalize the 2d crystal model to include the D4-branes on the compact 4-cycles of the Calabi-Yau, and then study the asymptotic analysis of these models. Presumably, this would produce the entropy of the dual black holes which are widely studied before, using other plausible techniques \cite{ma-st-wi,de-mo}. 

From the mathematical point of view, there are studies about the D4-D2-D0 brane and Donaldson-Thomas/Gromov-Witten invariants associated with the quiver gauge theory on the divisors of the Calabi-Yau \cite{bo-cr,gh-sh}. It would be interesting to investigate on the possible applications and interpretations of our asymptotic results in that context.

\section*{Acknowledgment}
I deeply thank A. Cazzaniga for the collaboration in early stages of this project. Thanks to Taro Kimura for interesting discussions and suggestions. The research of A.Z. has been supported by the French “Investissements d’Avenir” program, project ISITE-BFC (No. ANR-15-IDEX-0003), and EIPHI Graduate School (No. ANR-17-EURE-0002).

\bibliographystyle{plain}
\bibliography{references}

\end{document}